\documentclass[letterpaper, 10 pt, conference]{ieeetrans}  

\usepackage{mathptmx}
\usepackage[scaled=0.92]{helvet}
\usepackage{courier}
\usepackage{amsmath}
\usepackage{amssymb}
\usepackage{mathtools}
\usepackage{tikz}
\usepackage{algorithm}
\usepackage{algorithmic}
\usetikzlibrary{shapes,arrows}
\usepackage{tikz-3dplot}
\usetikzlibrary{arrows.meta,positioning}
\usetikzlibrary{positioning, shapes.multipart, calc}
\usepackage{url}
\usepackage{subcaption}
\usepackage{xcolor}

\newtheorem{definition}{Definition}

\newtheorem{remark}{Remark}

\newtheorem{proposition}{Proposition}

\newcommand{\bbR}{\mathbb{R}}

\IEEEoverridecommandlockouts                              

\title{\LARGE \bf
Learning and Verifying Stabilizing Controllers with Structured Neural Control Lyapunov Functions
}

\author{Jose Luis Ledesma Martin$^{1}$, Patricia Pauli$^{2}$
\thanks{$^{1}$Jose Luis Ledesma Martin is with the Department of Mechanical Engineering, Delft University of Technology, Delft, 2628 CD Delft, The Netherlands
        {\tt\small j.l.ledesmamartin@student.tudelft.nl}}%
\thanks{$^{2}$Patricia Pauli is with the Department of Mechanical Engineering, Eindhoven University of Technology, 5612 AZ Eindhoven, The Netherlands
        {\tt\small p.d.pauli@tue.nl}}%
}

\begin{document}

\maketitle
\thispagestyle{empty}
\pagestyle{empty}

\begin{abstract}
This work investigates the joint learning of neural controllers and control Lyapunov functions (CLFs) for nonlinear discrete-time systems. We exploit the structural properties of Input Convex Neural Networks and Polyak-Łojasiewicz Networks to guarantee positive definiteness of CLF candidates by construction, reducing the conditions requiring formal verification. We develop a mixed-integer programming framework to formally verify the remaining stability conditions and introduce an iterative region of attraction (ROA) expansion procedure to enlarge the certified ROA without retraining. Experiments on an inverted pendulum and a path-tracking system demonstrate the advantages of structured neural CLFs, particularly PLNets, and illustrate how their structural bounds influence the geometry of the certified regions. Our code is available at: \url{https://github.com/JLedesMart/Structured-Neural-Control-Lyapunov-Functions}.
\end{abstract}

\section{Introduction}

Neural networks provide expressive parameterizations of nonlinear functions and are therefore attractive for nonlinear controller synthesis, offering fast online evaluation without requiring online optimization. A well-established framework for stabilizing nonlinear systems is based on control Lyapunov functions (CLFs) \cite{chang2019neural}. A CLF is positive definite and certifies the existence of a feedback law under which the system decreases along the resulting closed-loop trajectories. In neural Lyapunov control, these conditions are typically enforced at sampled states during training and subsequently verified over a continuous region of the state space. Despite the flexibility of neural networks in representing both controllers and Lyapunov functions, their joint synthesis and formal certification remain challenging.

The joint synthesis of neural controllers and neural CLFs has been studied predominantly for continuous-time systems, often using counterexample-guided training together with satisfiability modulo theories (SMT)-based verification \cite{chang2019neural,zhou2022neural}. Recent work combines Zubov-based training with scalable neural-network verification to obtain less conservative certified inner approximations of the region of attraction (ROA) \cite{li2026two}. 
Related approaches construct formally verified neural CLFs from Hamilton-Jacobi-Bellman characterizations \cite{liu2025formally}. For discrete-time systems, \cite{dai2021lyapunov} proposed the joint synthesis of ReLU-based neural controllers and Lyapunov functions using counterexample-guided training and mixed-integer linear programming (MILP) verification for neural-network dynamics. Extending this approach to more general nonlinear dynamics, \cite{wu2023neural} incorporated certified bounds on the dynamics and gradient-based counterexample generation to improve training efficiency. More recent methods employ scalable bound propagation and branch-and-bound techniques for synthesis and verification \cite{yang2024lyapunov,shi2026certified}.



While existing approaches often encourage positive definiteness through explicit norm or quadratic terms, structured neural architectures provide CLF candidates that are positive definite by construction while retaining high expressive power. In particular, Input Convex Neural Networks (ICNNs) \cite{amos2017inputconvexneuralnetworks} are convex with respect to their inputs by construction and can be used to construct positive-definite CLF candidates \cite{kolter2019learning,Mukherjee2022}. Polyak-Łojasiewicz Networks (PLNets) \cite{wang2024monotone} satisfy prescribed lower and upper quadratic bounds, directly guaranteeing positive definiteness and coercivity. This work exploits these structural properties to develop expressive neural CLFs that satisfy the positivity conditions by construction, leaving only the closed-loop decrease condition to be established through formal verification.

Structured neural Lyapunov functions have primarily been used to enforce stability when learning unknown dynamics. For instance, \cite{cheng2024learningstablepassiveneural} parameterizes stable dynamics using a PLNet-based Lyapunov function, while \cite{kolter2019learning} projects learned dynamics onto a stability-preserving half-space defined by an ICNN-based Lyapunov function. Similarly, \cite{min2023data} jointly learns dynamics, controllers, and Lyapunov functions from data. Neural Lyapunov Differentiable Predictive Control \cite{Mukherjee2022} also employs ICNNs in discrete time, but optimizes control-input sequences with probabilistic guarantees. In contrast, the use of structured neural CLFs for synthesizing controllers with formal stability guarantees remains relatively unexplored, particularly for PLNets.

This paper develops a framework for jointly learning neural state-feedback controllers and CLFs for known nonlinear discrete-time systems, with formal stability verification over continuous regions of the state space. We consider ReLU controllers alongside PLNet, ICNN, and multilayer perceptron (MLP)-based CLF candidates, with positive definiteness ensured by construction for PLNets and ICNNs. Stability conditions are verified using MILP or mixed-integer quadratic programming (MIQP), and an iterative procedure expands the certified region of attraction (ROA).  Section \ref{sec:problem} formulates the problem and introduces the structured network architectures, Section \ref{sec:Training} presents the training procedure, Section \ref{sec:verification} develops the verification and ROA-expansion method, and Section \ref{sec:experiments} compares the certified ROAs on an inverted pendulum and a path-tracking system.

\section{Problem Statement}\label{sec:problem}
We consider the discrete-time dynamical system
\begin{equation}\label{eq:DT_system}
    x_{t+1} = f(x_t,u_t)
\end{equation}
with state vector $x_t\in X \subset \mathbb{R}^n$, control input $u_t\in\mathbb{R}^m$, and equilibrium point at the origin $x$ = 0. 
The stability of this system is characterized using a CLF.

\begin{definition}[Control Lyapunov Function] \label{def:clf} 
A function $V:\mathbb{R}^n\to\mathbb{R}$ is a control Lyapunov function for the discrete-time system \eqref{eq:DT_system} if 
\begin{enumerate} 
\renewcommand{\labelenumi}{(\roman{enumi})} 
\item $V(0)=0$, 
\item $V(x)>0$ $\forall x\neq 0$, 
\item $\forall x\neq 0,\ \exists u\in\mathbb{R}^{m} ~ \text{s.\,t.} ~ V\bigl(f(x,u)\bigr)-V(x)<0$. 
\end{enumerate}
\end{definition}

The goal of this work is to simultaneously find a valid neural CLF and a neural state-feedback controller that stabilizes the origin of the known system \eqref{eq:DT_system}. 



We parameterize CLF candidates using ReLU-based MLPs, ICNNs \cite{amos2017inputconvexneuralnetworks}, and PLNets \cite{wang2024monotone}, exploiting suitable constructions to guarantee conditions (i) and (ii) by design. ReLU-based MLPs are used as controllers, and condition (iii) is encouraged through joint training and subsequently verified formally.

\subsection{Multilayer Perceptron}  
\label{sec:mlp}
The MLP is a standard feedforward ReLU network given by
\begin{equation*}\label{eq:MLP}
\begin{aligned}
    z_k &= \mathrm{ReLU}(W_k z_{k-1} + b_k),
    \quad k=1,\ldots,L-1,\\
    g(x) &= W_L z_{L-1} + b_L,
\end{aligned}
\end{equation*}
where $z_0=x$, $L$ is the depth, and $W_k \in \mathbb{R}^{d_k \times d_{k-1}}$ and $b_k \in 
\mathbb{R}^{d_k}$ are the layer weights and biases. MLPs are used as controllers and as CLF candidates, with the latter parameterized as
\begin{equation}\label{eq:MLP_pos}
    V(x) = g(x)-g(0).
\end{equation}
We additionally consider a MLP-ReLU-L1
\begin{equation}\label{eq:MLP_ReLU_L1}
    V(x) = \mathrm{ReLU}(g(x)-g(0))
    + \epsilon\|x\|_1,\quad \epsilon>0,
\end{equation}
with $\epsilon > 0$ and $\|x\|_1$ denoting the L1 norm of the state, which guarantees conditions (i) and (ii) of
Definition~\ref{def:clf} by construction.

\subsection{Input convex neural networks}
\label{sec:icnn}
ICNNs \cite{amos2017inputconvexneuralnetworks} are convex with respect to their input and are recursively defined as
\begin{align*}
    z_k &= \mathrm{ReLU}(W_{k-1}^{(z)}z_{k-1}
    + W_{k-1}^{(x)}x + b_{k-1}), \quad k=1,\ldots,L,\\
    h(x) &= z_L,
\end{align*}
where $z_0=0$, $W_0^{(z)}=0$, and the matrices
$W_{k-1}^{(z)}$ have nonnegative entries to ensure convexity.

To guarantee conditions (i) and (ii) of Definition~\ref{def:clf}, we construct the CLF candidate, denoted as ICNN-L2, as
\begin{equation}\label{eq:ICNN-euclidean2}
    V(x) = \mathrm{ReLU}(h(x)-h(0)) + \epsilon\|x\|_2^2,
\end{equation}
where $\epsilon>0$ ensures positive definiteness and radial unboundedness through $V(x)\geq\epsilon\|x\|_2^2$. Alternatively, we consider
\begin{equation}\label{eq:ICNN-L1}
    V(x) = \mathrm{ReLU}(h(x)-h(0)) + \epsilon\|x\|_1,
\end{equation}
denoted as ICNN-L1, which preserves both properties while enabling MILP-based verification rather than MIQP-based verification, as discussed in Section~\ref{sec:verification}.

\subsection{Polyak-Łojasiewicz Network}
\label{sec:plnet}
The PLNet \cite{wang2024monotone}
is defined as
\begin{equation}\label{eq:PLNet_CLF}
    V(x)=\frac{1}{2}\|\mathcal{G}(x)-\mathcal{G}(0)\|_2^2,
\end{equation}
where $\mathcal{G}:\mathbb{R}^n\to\mathbb{R}^n$ is a
bi-Lipschitz network (BiLipNet) satisfying
\begin{equation}\label{eq:bilip}
    \mu\|x-x'\|_2
    \leq \|\mathcal{G}(x)-\mathcal{G}(x')\|_2
    \leq \nu\|x-x'\|_2,
\end{equation}
for all $x,x'\in\mathbb{R}^n$, with $0<\mu\leq\nu$.
The BiLipNet is constructed by composing orthogonal layers
$\mathcal{O}_k(x)=P_kx+q_k$, with $P_k^\top P_k=I$,
and feed-through network (FTN) layers $\mathcal{F}_k$
that are $\mu_k$-strongly monotone and $\nu_k$-Lipschitz:
\begin{equation*}
    \mathcal{G}
    =\mathcal{O}_K\circ\mathcal{F}_{K-1}
    \circ\cdots\circ\mathcal{F}_0\circ\mathcal{O}_0.
\end{equation*}
Each FTN layer is of the form
\begin{align*}
    z_j &= \mathrm{ReLU}(W_jz_{j-1}+U_jx+b_j),
    \quad j=1,\ldots,L,\\
    \mathcal{F}_k(x) &= \mu_k x+\sum_{j=1}^L Y_jz_j+b_y,
\end{align*}
with $z_0=0$ and trainable parameters $U_j$, $W_j$, $Y_j$, $b_j$, $b_y$ of appropriate
dimensions. The parameterization in
\cite{wang2024monotone}
guarantees the required monotonicity and Lipschitz bounds
by construction. Consequently, $\mathcal{G}$ is
bi-Lipschitz with
$\mu=\prod_{k=0}^{K-1}\mu_k$ and
$\nu=\prod_{k=0}^{K-1}\nu_k$ and the corresponding PLNet \eqref{eq:PLNet_CLF} satisfies
\begin{equation*}
    \frac{\mu^2}{2}\|x\|_2^2
    \leq V(x)
    \leq \frac{\nu^2}{2}\|x\|_2^2.
\end{equation*}
Unlike the ICNN-based construction, which guarantees only
a quadratic lower bound, the PLNet provides both lower
and upper bounds. The ratio $\mu:\nu$ constrains the geometry of the Lyapunov sublevel sets and thereby influences the shape of the certified ROA approximation, as discussed in Section~\ref{sec:shape_ROA}.

\section{Training}\label{sec:Training}
Our objective is to train a neural CLF candidate and a neural controller using backpropagation and gradient-based optimization. In Subsection~\ref{sec:costfunction}, we first introduce the Lyapunov risk used as the training loss, followed by the joint training procedure in Subsection~\ref{sec:Lyap_controller}.

\subsection{Lyapunov Risk}\label{sec:costfunction}
The Lyapunov risk jointly optimizes the neural CLF candidate
$V_\phi$ and controller $\pi_\theta$ by penalizing violations
of conditions (ii) and (iii) of Definition~\ref{def:clf}.
Since all considered CLF architectures satisfy $V_\phi(0)=0$
by construction, condition (i) requires no loss term.
The Lyapunov risk reads
\begin{align}\label{eq:Lyap_risk}
    \mathcal{L}_{\mathrm{lyap}}
    &= \mathcal{L}_{\text{(ii)}} + \mathcal{L}_{\text{(iii)}},\\
    \mathcal{L}_{\text{(ii)}}
    &= \beta\,\mathrm{ReLU}\bigl(
    -V_\phi(x) + \rho\min(\|x\|_2,\kappa)\bigr),\nonumber\\
    \mathcal{L}_{\text{(iii)}}
    &= \mathrm{ReLU}\bigl(
    V_\phi(f(x,\pi_\theta(x))) - V_\phi(x) + \delta\bigr).\nonumber
\end{align}
Here, $\beta > 0$ weights the penalty for violating condition~(ii), while $\delta \geq 0$ specifies the desired decrease margin in condition~(iii). The hyperparameters $\rho>0$ and $\kappa>0$ define a positivity margin that encourages $V_\phi$ to increase with the Euclidean norm near the origin, thereby discouraging flat Lyapunov functions.

When a MLP-ReLU-L1~\eqref{eq:MLP_ReLU_L1}, an ICNN \eqref{eq:ICNN-euclidean2} or \eqref{eq:ICNN-L1}, or a PLNet \eqref{eq:PLNet_CLF} is used as the CLF candidate, condition (ii) of Definition~\ref{def:clf} is satisfied by construction. Consequently, only the decrease condition (iii) must be enforced during training, and the Lyapunov risk reduces to $\mathcal{L}_{lyap} = \mathcal{L}_{\text{(iii)}}$.

\subsection{Learning neural CLF and neural controller functions}\label{sec:Lyap_controller}
The following training strategy is based on \cite{wu2023neural}. 
Joint training of the neural controller and CLF from random initialization often fails to converge. Although the PLNet or ICNN architecture yield a well-structured CLF candidate that satisfies conditions (i) and (ii) by construction, the randomly initialized MLP controller is highly nonlinear and may induce unstable closed-loop dynamics, thereby impeding training.
To achieve successful training, we follow a multi-phase approach, showcased in Algorithm \ref{alg:training}:\\
\textbf{Phase 1:} The neural controller is first trained to approximate a linear quadratic regulator (LQR) controller designed for the system dynamics \eqref{eq:DT_system} linearized about the equilibrium.\\
\textbf{Phase 2:} The CLF candidate is trained to attempt to satisfy the Lyapunov conditions (i) to (iii) of Definition \ref{def:clf} for this controller so that it learns a reasonable initial shape. \\
\textbf{Phase 3:} The neural controller and the CLF are trained together to satisfy the Lyapunov conditions (i) to (iii) of Definition \ref{def:clf} and are submitted to the formal verifier. \\

\begin{algorithm}[h!]
\caption{Training Control and Lyapunov Networks}
\label{alg:training}
\begin{algorithmic}[1]
\REQUIRE System dynamics $f$, domain $X$, training Epochs $E_1, E_2, E_3$
\ENSURE Certified neural controller $\pi_\theta$, CLF candidate $V_\phi$
\STATE Calculate LQR gain matrix $K$ from linearized dynamics

\STATE \COMMENT{\textbf{Phase 1: Controller Imitation Learning}}
\FOR{epoch = 1 \TO $E_1$}
    \STATE Sample states $x \sim X$
    \STATE Compute LQR target control: $u^*$
    \STATE $\mathcal{L}_{\text{MSE}} = ||\pi_\theta(x) - u^*||^2$
    \STATE $\theta \leftarrow \theta - \alpha \nabla_\theta\mathcal{L}_{\text{MSE}} $
\ENDFOR

\STATE \COMMENT{\textbf{Phase 2: Lyapunov Network Warm-up}}
\STATE Initialize buffer $B$ with $N$ points sampled uniformly
from an enlarged training region $\widetilde{X}\supset X$
\STATE $S \leftarrow B$
\FOR{epoch = 1 \TO $E_2$}
    \IF{epoch $\mod 10 = 0$}
        \STATE Sample r points from $B$: $\hat{S} = \text{Sample}(r,B)$
        \STATE Sample q points from $\widetilde{X}\supset X$: $R = \text{Sample}(q,\widetilde{X})$
        \STATE $F \leftarrow$ \textsc{FindCounterExamples}$(R)$
        \STATE $B \leftarrow F \cup B$ 
        \STATE $S \leftarrow \hat{S} \cup F$
        
    \ENDIF
    \STATE $\phi \leftarrow \phi - \alpha \nabla_\phi \mathcal{L}_{\text{lyap}}(V_\phi, S)$
\ENDFOR

\STATE \COMMENT{\textbf{Phase 3: Joint Training}}
\STATE Verified = False
\STATE $C \leftarrow \varnothing$
\WHILE{\textbf{not} Verified}
\STATE Initialise buffer $B$ with $N$ points sampled uniformly from $\widetilde{X}\supset X$
\STATE $S \leftarrow B \cup C$
\FOR{epoch = 1 \TO $E_3$}
    
    \IF{epoch $\mod 10 = 0$}
        \STATE Sample r points from $B$: $\hat{S} = \text{Sample}(r,B)$
        \STATE Sample q points from $\widetilde{X}\supset X$: $R = \text{Sample}(q,\widetilde{X})$
        \STATE $F \leftarrow$ \textsc{FindCounterExamples}$(R)$
        \STATE $B \leftarrow F \cup B$ 
        \STATE $S \leftarrow \hat{S} \cup F \cup C$
        
    \ENDIF
    \STATE $\theta \leftarrow \theta - \alpha \nabla_{\theta} \mathcal{L}_{\text{lyap}}(V_\phi, \pi_\theta, S)$
    \STATE $\phi \leftarrow \phi - \alpha \nabla_{\phi} \mathcal{L}_{\text{lyap}}(V_\phi, \pi_\theta, S)$
\ENDFOR
\STATE Verified, $\hat{C}$ = \textsc{Certify}($\pi_\theta$,$V_\phi$)
\STATE$C \leftarrow \hat{C} \cup C$
\ENDWHILE
\end{algorithmic}
\end{algorithm}

In Phases 2 and 3, we employ a \textsc{FindCounterExamples} function following \cite{wu2023neural}, which uses projected gradient descent (PGD) to efficiently identify states violating the Lyapunov conditions. For PLNet, ICNN, and MLP-ReLU-L1-based CLFs, this involves finding points that violate the decrease condition~(iii) by solving
\begin{equation*}
\max_{x\in X} V_\phi(f(x, \pi_\theta(x))) -V_\phi(x).
\end{equation*}
Starting from randomly sampled states, we iteratively search for worst-case violations and project each update back onto the training region $\widetilde{X}\supset X\in\bbR^n$, $X$ denoting the domain. The returned set $F$ contains only states that violate condition (iii). 


\section{Verification}\label{sec:verification}
If the training loss \eqref{eq:Lyap_risk} attains zero, conditions (i) to (iii) are satisfied on all training samples. These conditions must subsequently be verified over the entire domain of interest. 

We build on the work in \cite{wu2023neural}, where an MILP-based verification is used, extending their approach to ICNNs and PLNets. When the CLF candidate contains quadratic terms, such as the PLNet construction \eqref{eq:PLNet_CLF} or a quadratic regularization term $\epsilon\|x\|_2^2$, the verification problem has a quadratic objective instead of a linear one in , resulting in a generally nonconvex MIQP. The resulting quadratic objective generally increases computational complexity and verification time. 
Following \cite{wu2023neural}, we utilize the IBM ILOG CPLEX optimization suite \cite{IBMCPLEX}.

\begin{remark}[Verification near the equilibrium]
\label{rem:epsilon_stability}
Verifying strict Lyapunov decrease arbitrarily close to the
equilibrium is numerically challenging, as the decrease
approaches zero. Following \cite{wu2023neural}, we
therefore exclude a small neighborhood $B_\varepsilon$
of the origin from verification. Under suitable Lipschitz
continuity, invariance, and equilibrium conditions, this
can provide practical stability guarantees
\cite{wu2023neural}, but does not establish asymptotic
stability of the origin.
\end{remark}


\subsection{MILP and MIQP formulation} \label{subsec:MILPMIQP}
The domain $X$ is partitioned into $N_c$ cells
$\mathcal{C}_i$ 
\begin{equation}
    X = \bigcup_{i=1}^{N_c} \mathcal{C}_i,
    \quad
    \operatorname{int}(\mathcal{C}_i)
    \cap
    \operatorname{int}(\mathcal{C}_j)
    = \emptyset
    \quad \forall i \neq j,
\end{equation}
where each cell is a hyperrectangle of the form
\begin{equation}
    \mathcal{C}_i =
    \{x \in \mathbb{R}^n \mid l_i \leq x \leq u_i\},
    \quad i=1,\ldots,N_c,
\end{equation}
with $l_i,u_i\in\mathbb{R}^n$ denoting the componentwise
lower and upper bounds of cell $\mathcal{C}_i$. The partition is problem-dependent. In our experiments, we use the same cells as \cite{wu2023neural}. 

The CLF decrease condition (iii) in Definition~\ref{def:clf} is verified by solving, for each cell $\mathcal{C}_i$, the optimization problem
\begin{equation}\label{eq:verification}
\begin{aligned}
\gamma_{\mathcal{C}_i}^\star = \max_{x_t,u_t,x_{t+1}} \quad
& V_\phi(x_{t+1}) - V_\phi(x_t) \\
\text{subject to} \quad
& x_t \in \mathcal{C}_i,\\
& u_t = \pi_\theta(x_t),\\
& x_{t+1} = f(x_t,u_t).
\end{aligned}
\end{equation}
If $\gamma_{\mathcal{C}_i}^\star<0$ the cell $\mathcal{C}_i$ is verified. If verification succeeds for every cell $i=1,\dots,N_c$, the entire region $X$ is certified. Note that for MLP-based CLF candidates \eqref{eq:MLP_pos} condition (ii) is verified on every cell through a similar optimization problem. 

To obtain a tractable relaxation of \eqref{eq:verification}, we encode the controller and CLF candidate, evaluating the latter at both $x_t$ and $x_{t+1}$. This yields an MILP or an MIQP for PLNets~\eqref{eq:PLNet_CLF} and ICNN-L2~\eqref{eq:ICNN-euclidean2} due to their quadratic output structure. The resulting optimization problem provides an upper bound $\overline{\gamma}_{\mathcal{C}_i}$ on the optimal value of the original verification problem, i.e., $\gamma_{\mathcal{C}_i}^\star \leq \overline{\gamma}_{\mathcal{C}_i}$. Consequently, $\overline{\gamma}_{\mathcal{C}_i}<0$ certifies the Lyapunov decrease condition (iii) of Definition \ref{def:clf} throughout $\mathcal{C}_i$.  

Following \cite{wu2023neural} and the discussion in Remark~\ref{rem:epsilon_stability}, we exclude a small neighborhood of the equilibrium from verification. The network encodings and affine relaxations of the dynamics are detailed in the following subsections.

\subsection{Bounding and encoding the dynamics}
To represent the nonlinear dynamics $x_{t+1}=f(x_t,u_t)$ using linear constraints, certified affine lower and upper bounds on $f$ are computed over each cell $\mathcal C_i$. Following \cite{wu2023neural}, we obtain these bounds over the corresponding input domain using $\alpha,\beta$-CROWN \cite{xu2020automatic}, such that
\begin{equation*}
\underline{f}_i(x_t,u_t)
\leq
f(x_t,u_t)
\leq
\overline{f}_i(x_t,u_t),
\end{equation*}
where $\underline{f}_i$ and $\overline{f}_i$ are affine functions and the inequalities hold componentwise. These bounds 
provide an outer approximation of the nonlinear dynamics over $\mathcal C_i$.


\subsection{Bounding and encoding neural networks}
We use interval bound propagation (IBP) \cite{gowal2019scalable} to compute componentwise
bounds on the pre- and post-activations of the neural networks
over each input cell $\mathcal{C}_i$. For an affine layer
$\hat{z}_k=W_k z_{k-1}+b_k$, let
$\gamma_{k-1}$ and $r_{k-1}$ denote the center and
radius of the input interval. The corresponding
pre-activation interval is characterized by
\begin{equation*}
    \hat{\gamma}_k=W_k\gamma_{k-1}+b_k,
    \qquad
    \hat{r}_k=|W_k|r_{k-1},
\end{equation*}
where $|W_k|$ denotes the elementwise absolute value.
The pre-activation bounds follow as
$\underline{\hat{z}}_k=\hat{\gamma}_k-\hat{r}_k$ and
$\overline{\hat{z}}_k=\hat{\gamma}_k+\hat{r}_k$.
Post-activation bounds are obtained by applying ReLU
to both bounds componentwise. For ICNNs and BiLipNets,
the same procedure applies, accounting for their additional
input connections and layer structures, specified in Subsections \ref{sec:icnn} and \ref{sec:plnet}, respectively.

The resulting bounds are used to encode each ReLU exactly
using mixed-integer linear constraints. For neurons with
$\underline{\hat{z}}_{k,j}<0<\overline{\hat{z}}_{k,j}$,
we introduce a binary variable $a_{k,j}\in\{0,1\}$
and an auxiliary variable $s_{k,j}\geq0$, yielding
\begin{equation}\label{eq:ReLU_encoding}
\begin{aligned}
\hat{z}_{k,j}
&=z_{k,j}-s_{k,j},\\
0\leq z_{k,j}
&\leq\overline{\hat{z}}_{k,j}a_{k,j},\\
0\leq s_{k,j}
&\leq-\underline{\hat{z}}_{k,j}(1-a_{k,j}).
\end{aligned}
\end{equation}
Neurons with fixed activation states require no binary variables:
if $\underline{\hat{z}}_{k,j}\geq0$, then
$z_{k,j}=\hat{z}_{k,j}$, whereas if
$\overline{\hat{z}}_{k,j}\leq0$, then $z_{k,j}=0$.

\subsection{Computing certified ROAs}
\label{subsec:ROA}

We seek a large certified inner approximation of the ROA.
\begin{definition}[Certified Region of Attraction]
\label{def:roa}
A set 
\begin{equation*}
R_c=\{x\in\mathbb{R}^n\mid V_\phi(x)\leq c\},
\end{equation*}
with $c>0$, is a certified inner approximation of the
ROA if all trajectories initialized
in $R_c$ are guaranteed to converge to the equilibrium
under the closed-loop dynamics.
\end{definition}


Following \cite{wu2023neural}, we determine a certified ROA sublevel set $R_c$ (Definition~\ref{def:roa}) contained in the verified region $X$ using Proposition~\ref{prop:roa}.

\begin{proposition}[Certified ROA Sublevel Set]
\label{prop:roa}
Let the closed-loop system satisfy \(f(0,\pi_\theta(0))=0\), and let \(V_\phi\) be continuous and positive definite. Suppose that \(V_\phi(f(x,\pi_\theta(x)))-V_\phi(x)<0\) for all \(x\in X\setminus\{0\}\), and let \( X_{\mathrm{outer}}\supset X\) contain all one-step successors of states in \(X\). 
Define
\begin{equation}
\label{eq:B_and_cstar}
B=X_{\mathrm{outer}}\setminus X,
\qquad
c^*=\min_{x\in B}V_\phi(x).
\end{equation}
Then, for any $0<c<c^*$, the sublevel set $R_c$
defined in Definition~\ref{def:roa} is contained in
$X$, is forward invariant, and constitutes a certified
inner approximation of the ROA.
\end{proposition}

\begin{proof}
Let $x_t\in R_c$. Since $R_c\subseteq X$, for all $x_t\in R_c\setminus\{0\}$, the Lyapunov decrease
condition (iii) implies
\begin{equation*}
V_\phi(x_{t+1})<V_\phi(x_t)\leq c<c^*.
\end{equation*}

Suppose, for contradiction, that $x_{t+1}\notin X$. Since
$x_{t+1}\in X_{\mathrm{outer}}$ by assumption, we have
$x_{t+1}\in B$, and hence $V_\phi(x_{t+1})\geq c^*$, contradicting the preceding inequality. Thus, $x_{t+1}\in X$.
Together with $V_\phi(x_{t+1})\leq c$, this implies
$x_{t+1}\in R_c$, establishing forward invariance.

Since the Lyapunov conditions hold on $X$ and trajectories
initialized in $R_c$ remain in $R_c\subseteq X$, they converge
to the equilibrium. Therefore, $R_c$ is a certified inner
approximation of the ROA.
\end{proof}

The certified and outer regions are chosen as
\begin{align*}
X
&=\{x\in\mathbb{R}^n\mid \|x\|_\infty<d\},\\
X_{\mathrm{outer}}
&={\{x\in\mathbb{R}^n\mid \|x\|_\infty\leq d+\Delta d\}},
\end{align*}
where $\Delta d$ is selected such that all one-step successors of states in
$X$ are contained in $X_{\mathrm{outer}}$. Since $V_\phi$ is continuous and
positive definite and $B$ is compact with $0\notin B$, the minimum in
\eqref{eq:B_and_cstar} exists and satisfies $c^*>0$. In practice, we choose
$c=c^*-\epsilon$ for a small $\epsilon>0$.

\subsection{Expanded ROA} \label{sec:expandedROA}
\begin{figure}[h]
    \centering
    \includegraphics[width=0.8\linewidth]{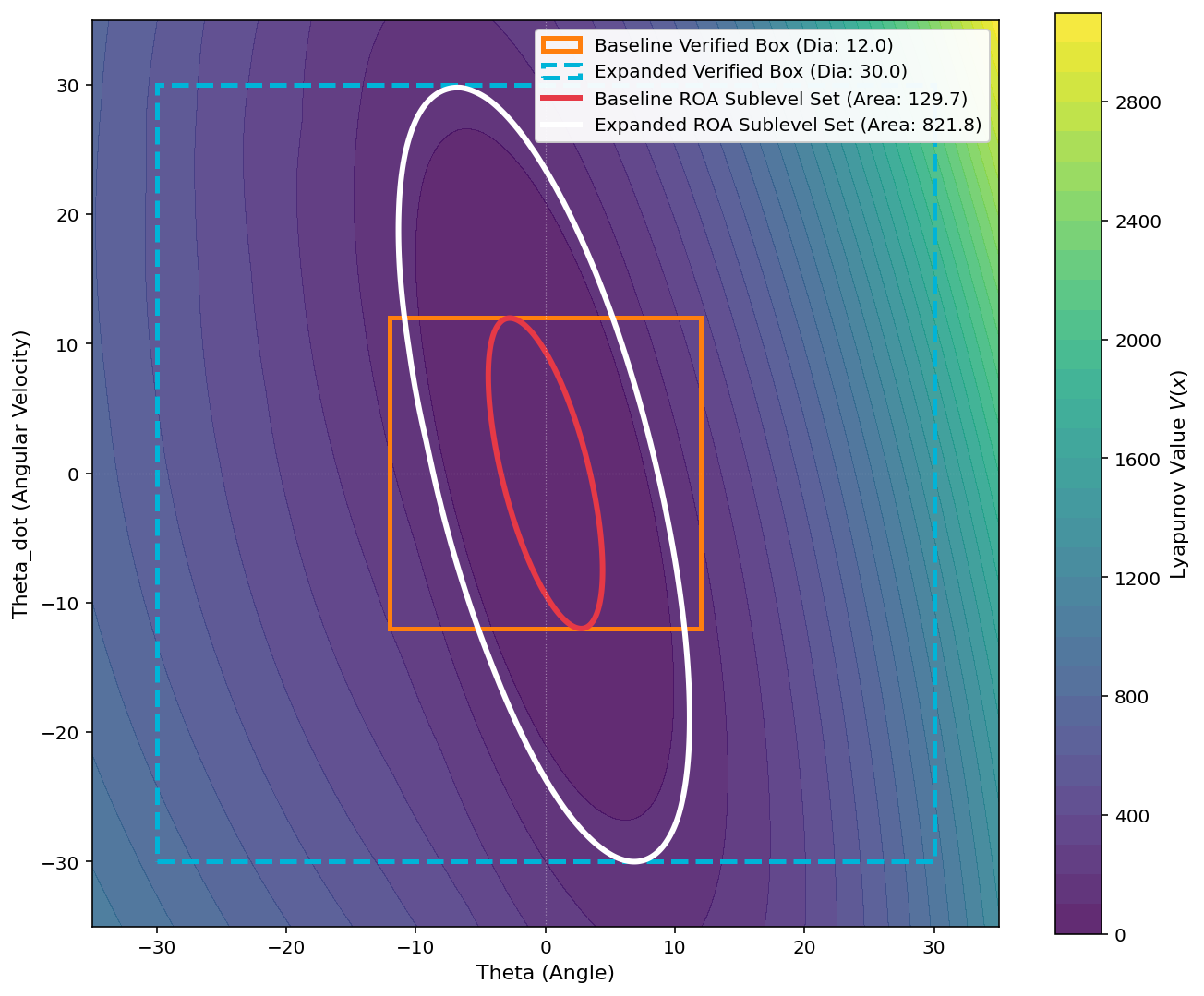}
    \caption{Result of the ROA expansion procedure. The certified region is incrementally expanded by verifying rings around the original training region $X$, yielding a larger certified ROA}
    \label{fig:roa_expansion}
\end{figure}

Following \cite{wu2023neural}, an initial ROA estimate is obtained once the MILP/MIQP verifier certifies that conditions (i) to (iii) of Definition~\ref{def:clf} hold throughout the training region $X$. However, this estimate may be conservative, as the true ROA can extend beyond $X$. To obtain a less conservative estimate, we propose an iterative
expansion procedure that incrementally enlarges the
verified domain. Specifically, the domain size
parameter $d$ is increased by a fixed step size,
and the Lyapunov conditions are verified over
successive rings
$\{x \mid d_k \leq \|x\|_\infty \leq d_{k+1}\}$.
The procedure terminates when verification fails,
and the largest successfully verified domain
replaces $X$ in Proposition~\ref{prop:roa}
to compute an expanded ROA estimate.
This procedure requires no retraining of the networks
and is detailed in Algorithm~\ref{alg:roa_expansion},
with an illustration in Fig.~\ref{fig:roa_expansion}.
\begin{algorithm}[h!]
\caption{ROA Expansion}
\label{alg:roa_expansion}
\begin{algorithmic}[1]
\REQUIRE Policy $\pi_\theta$, CLF $V_\phi$, base diameter $d_0$, step size $\Delta$, maximum diameter $d_{\max}$
\ENSURE Expanded ROA
\STATE $d_{\text{last}} \leftarrow d_0$
\STATE $d \leftarrow d_0 + \Delta$
\WHILE{$d \leq d_{\max}$}
    \STATE Verify ring $\{x \mid d - \Delta \leq \|x\|_\infty \leq d\}$ using MILP/MIQP
    \IF{ring certified}
        \STATE $d_{\text{last}} \leftarrow d$
        \STATE $d \leftarrow d + \Delta$
    \ELSE
        \STATE \textbf{break}
    \ENDIF
\ENDWHILE
\STATE Let $X^* = \{x \mid \|x\|_\infty \leq d_{\text{last}}\}$ be the largest certified region
\STATE Let $X^*_{\text{outer}} = \{x \mid \|x\|_\infty \leq d_{\text{last}} + \Delta d\}$ where $\Delta d$ is the maximum single-step displacement
\STATE $c^* \leftarrow \min_{x \in X^*_{\text{outer}} \setminus X^*} V_\phi(x)$
\RETURN $R^* = \{x \in X^* \mid V_\phi(x) \leq c^* - \epsilon\}$
\end{algorithmic}
\end{algorithm}

\begin{remark}
The proposed ROA expansion algorithm demonstrates that the certified ROA can be extended beyond the initial training region. While we employ a simple expansion strategy with fixed step sizes, more sophisticated strategies, such as adaptive or nonuniform expansion, may yield even larger certified ROA estimates.
\end{remark}

\section{Experiments}\label{sec:experiments}

\subsection{Experimental setup}
We evaluate the proposed approach on two nonlinear dynamical systems: an inverted pendulum and a path-tracking system, both considered in \cite{wu2023neural}. For both systems, the nonlinear dynamics are discretized using forward Euler integration with a sampling time of $\Delta t=0.05$ s. The dynamics are also linearized around the equilibrium, and the resulting linear systems are discretized with the same sampling time. The LQR gain matrix $K$ is computed by solving the discrete-time algebraic Riccati equation with $Q=I$ and $R=10$ using the Python control package, providing the reference controller for Phase 1 of Algorithm~\ref{alg:training}.

\subsubsection{Inverted Pendulum} The continuous-time non-linear dynamics of the inverted pendulum are described as
\begin{equation*}
    \ddot{\theta} = \frac{mgl \sin{\theta} + \tau - 0.1 \dot{\theta}}{ml^2}.
\end{equation*}
The state is $x=\begin{bmatrix}\theta & \dot{\theta} \end{bmatrix}^\top$, with control torque $\tau$ constrained by $|\tau|\leq\tau_{\max}$. The dynamics can be written as \eqref{eq:DT_system} with
\begin{equation*}
    f(x, u) = \begin{bmatrix} \dot{\theta} \\ \dfrac{mgl\sin\theta + \tau - 0.1\dot{\theta}}{ml^2} \end{bmatrix}.
\end{equation*} The equilibrium is $(x^*,u^*)=(0,0)$, yielding the LQR controller $u_{\mathrm{LQR}}=-Kx$. Following \cite{wu2023neural}, the training region is $\|x\|_\infty\leq 12$, and the ROA expansion procedure described in Algorithm \ref{alg:roa_expansion} is performed up to $\|x\|_\infty\leq 30$.

\begin{table}[t]
\centering
\setlength{\tabcolsep}{3pt}

\begin{minipage}[t]{0.48\linewidth}
\centering
\begin{tabular}{lll}
\hline
Parameter & Symbol & Value \\
\hline
Mass & $m$ & 0.15 kg \\
Length & $l$ & 0.5 m \\
Gravity & $g$ & 9.81 m/s$^2$ \\
Max torque & $\tau_{\max}$ & 6.0 Nm \\
Timestep & $\Delta t$ & 0.05 s \\
\hline
\end{tabular}
\caption{Inverted pendulum parameters.}
\label{tab:pendulum_params}
\end{minipage}
\hfill
\begin{minipage}[t]{0.48\linewidth}
\centering
\begin{tabular}{lll}
\hline
Parameter & Symbol & Value \\
\hline
Speed & $v$ & 2.0 m/s \\
Length & $l$ & 1.0 m \\
Radius & $r$ & 10.0 m \\
Max steering & $u_{\max}$ & 0.84 \\
Timestep & $\Delta t$ & 0.05 s \\
\hline
\end{tabular}
\caption{Path tracking parameters.}
\label{tab:path_params}
\end{minipage}

\end{table}

\subsubsection{Path Tracking} 
Following \cite{Snider2009AutomaticSM,chang2019neural,wu2023neural}, the continuous-time kinematic path-tracking dynamics are given by
\begin{equation*}
    f(x, u) = \begin{bmatrix} v\sin\theta_e \\ \dfrac{vu}{l} - \dfrac{v\cos\theta_e}{r - d_e} \end{bmatrix}.
\end{equation*}
The state $x=\begin{bmatrix} d_e & \theta_e \end{bmatrix}^\top$ represents the radial and angular deviations from a circular reference path of radius $r$, followed at constant speed $v$. The control input is $u=\tan(\delta)$, with steering angle $\delta\in[-40^\circ,40^\circ]$. The equilibrium is $(x^*,u^*)=(0,l/r)$, yielding $u_{\mathrm{LQR}}=l/r-Kx$. Following \cite{wu2023neural}, the training region is $\|x\|_\infty\leq 3$, with ROA expansion up to $\|x\|_\infty\leq 9$.

\subsubsection{Neural architectures and training}
For each benchmark, we jointly train a neural controller and a CLF candidate using six parameterizations: MLP, MLP-ReLU-L1, ICNN-L2, ICNN-L1, PLNet, defined in \eqref{eq:MLP_pos}, \eqref{eq:MLP_ReLU_L1}, \eqref{eq:ICNN-euclidean2}, \eqref{eq:ICNN-L1}, and \eqref{eq:PLNet_CLF}, respectively, and a quadratic baseline $V(x)=x^\top Px$, $P\succ0$. The neural CLF architectures have comparable numbers of trainable parameters (342 for PLNet, 323 for ICNN, and 321 for MLP), while all controllers use the same MLP architecture with 177 parameters. For PLNets, we set $\mu=0.5$ and $\nu=2$. Each architecture is evaluated over five independent runs (seeds 11–15). In Phase 1, the controller is first trained to imitate the LQR controller for 1000 epochs with learning rate $10^{-3}$, followed by 500 epochs each for Phases 2 and 3 with learning rate $10^{-2}$.



\subsubsection{Verification}
The ICNN-L2, PLNet, and quadratic CLF candidates are verified using MIQP, whereas the ICNN-L1, MLP, and MLP-ReLU-L1 candidates are verified using MILP. Certification is performed using CPLEX 22.2.0.0, with a 15-minute time limit per run. A run is considered unsuccessful if the initial training region cannot be certified within this limit. All experiments are conducted on an AMD Ryzen 7 5800X processor with 32 GB RAM, using 14 parallel verification workers.


\subsubsection{Evaluation metrics}
We compare the architectures using five evaluation metrics:\\
\textbf{ROA:} Area of the certified sublevel set
    (Section~\ref{subsec:ROA}), reported before expansion
    within the training domain $X$ (base ROA) and after
    applying Algorithm~\ref{alg:roa_expansion} (expanded ROA).\\
\textbf{Certified region diameter:} The value
    $d_{\mathrm{last}}$ defining the largest verified region
    $\{x \mid \|x\|_\infty \leq d_{\mathrm{last}}\}$,
    obtained using Algorithm~\ref{alg:roa_expansion}.\\
\textbf{Runtime:} Wall-clock time, reported
    separately for Phases 2 and 3, ROA expansion, and
    the total. Experiments are executed sequentially.
    For the quadratic CLF, only the controller is trained,
    so the Phase 2 and 3 times are combined.\\
\textbf{Certification rounds:} Number of iterations
    of the verification loop in Phase 3 of
    Algorithm~\ref{alg:training} (lines 25--41).\\
\textbf{Success rate:} Percentage of runs in which
    the initial training domain is successfully certified
    within the prescribed time limit.



All metrics are reported as mean $\pm$ standard deviation over five seeds, with the maximum base and expanded ROAs reported additionally. For failed runs, the base and expanded ROAs and expanded region diameter are set to zero and included in the statistics. Phase 3 runtime, expansion runtime, total runtime, and certification rounds are averaged over successful runs only, whereas Phase 2 runtime is averaged over all runs, as this phase always completes.


\subsection{Comparison of CLF Architectures}

\begin{figure}[h]
    \centering
    \begin{subfigure}[b]{0.23\textwidth}
        \centering
        \includegraphics[width=\textwidth]{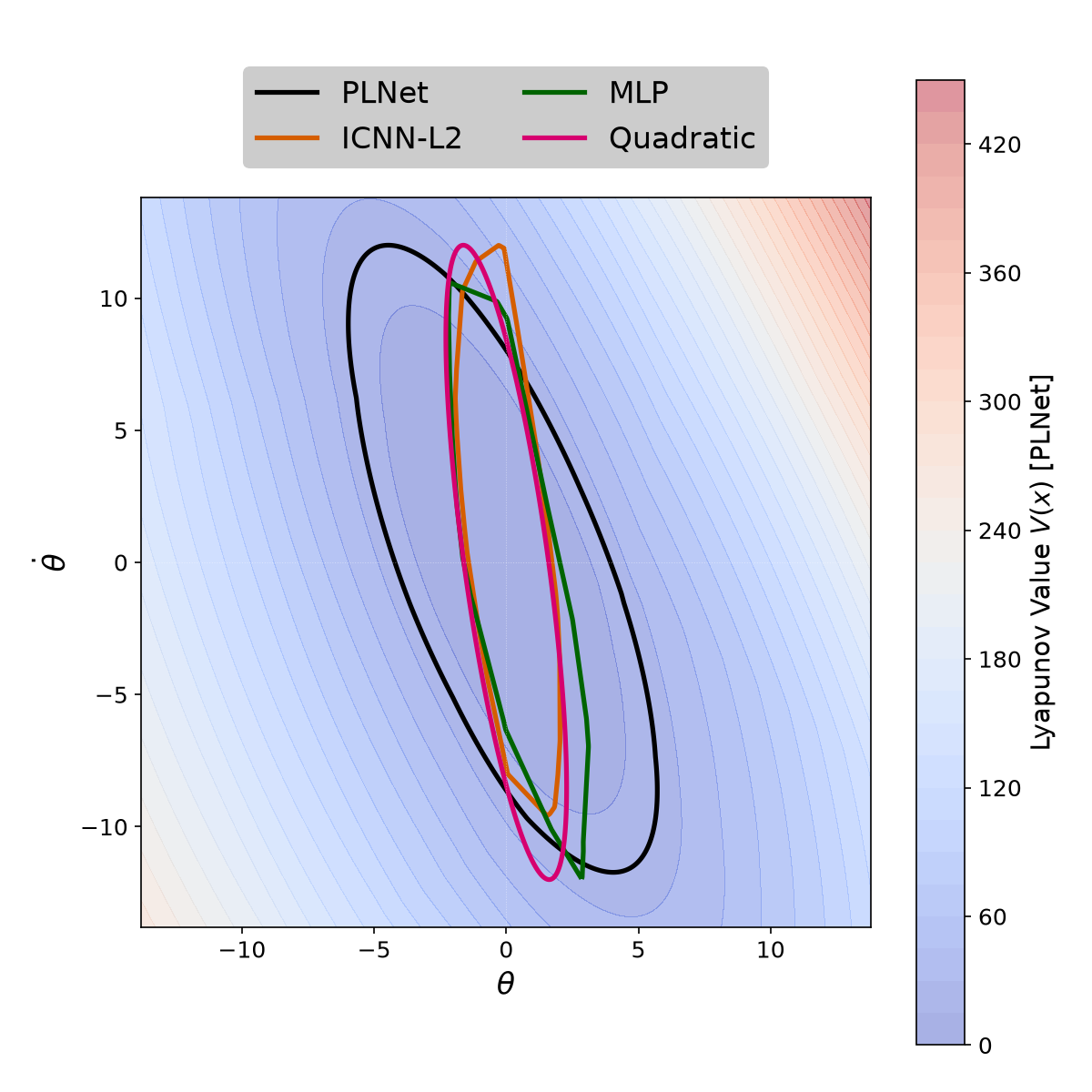}
        \caption{Baseline ROA sublevel sets ($V(x) \leq c_{\text{train}}$).}
        \label{fig:IPbase}
    \end{subfigure}
    \begin{subfigure}[b]{0.23\textwidth}
        \centering
        \includegraphics[width=\textwidth]{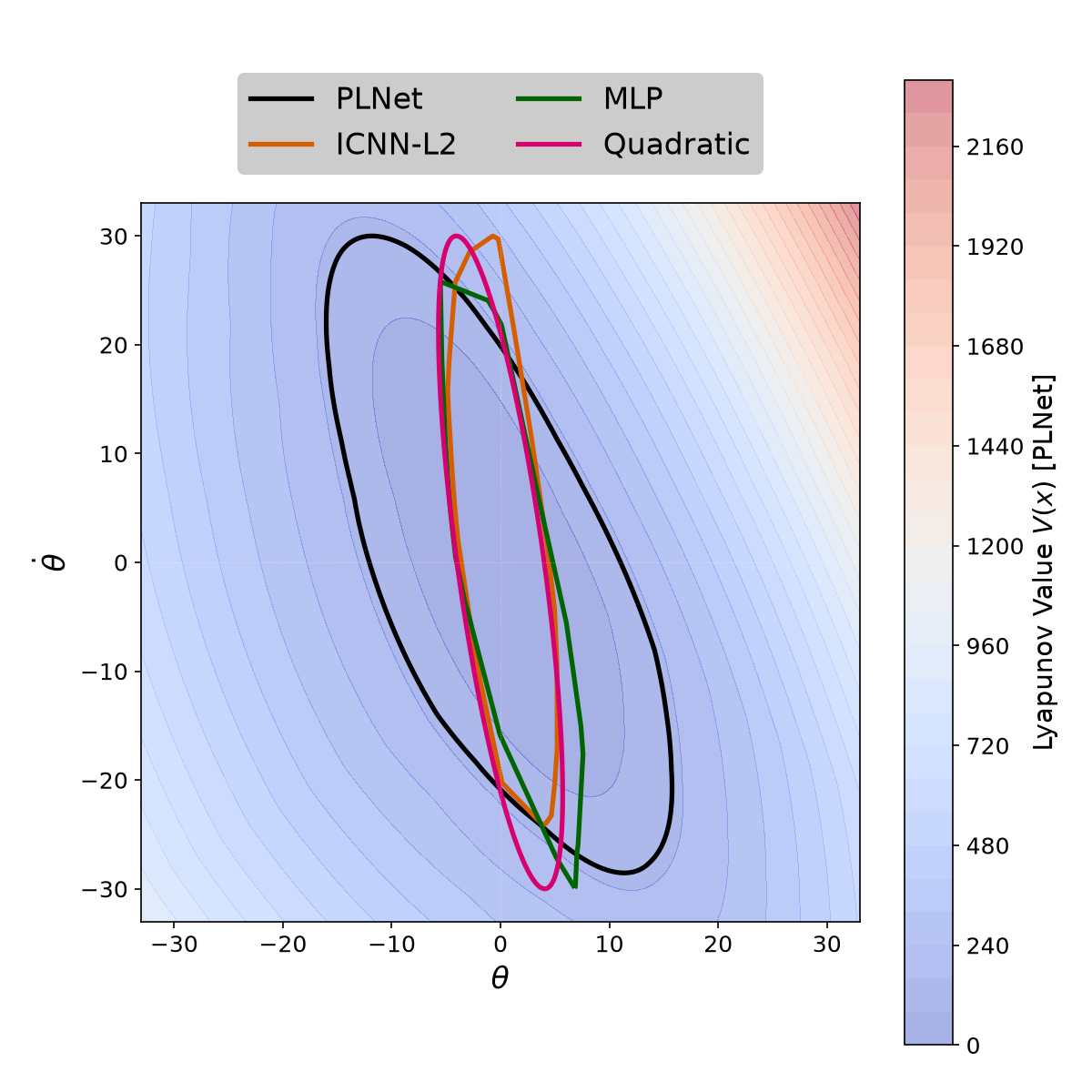}
        \caption{Expanded ROA sublevel sets ($V(x) \leq c_{\text{expanded}}$).}
        \label{fig:IPexpanded}
    \end{subfigure}
    \caption{Comparison of certified ROA for the inverted pendulum across PLNet, ICNN-L2, MLP and quadratic CLF architectures, showing the best performing seed for each.}
    \label{fig:ArchIP}
\end{figure}

\begin{figure}[h]
    \centering
    \begin{subfigure}[b]{0.23\textwidth}
        \centering
        \includegraphics[width=\textwidth]{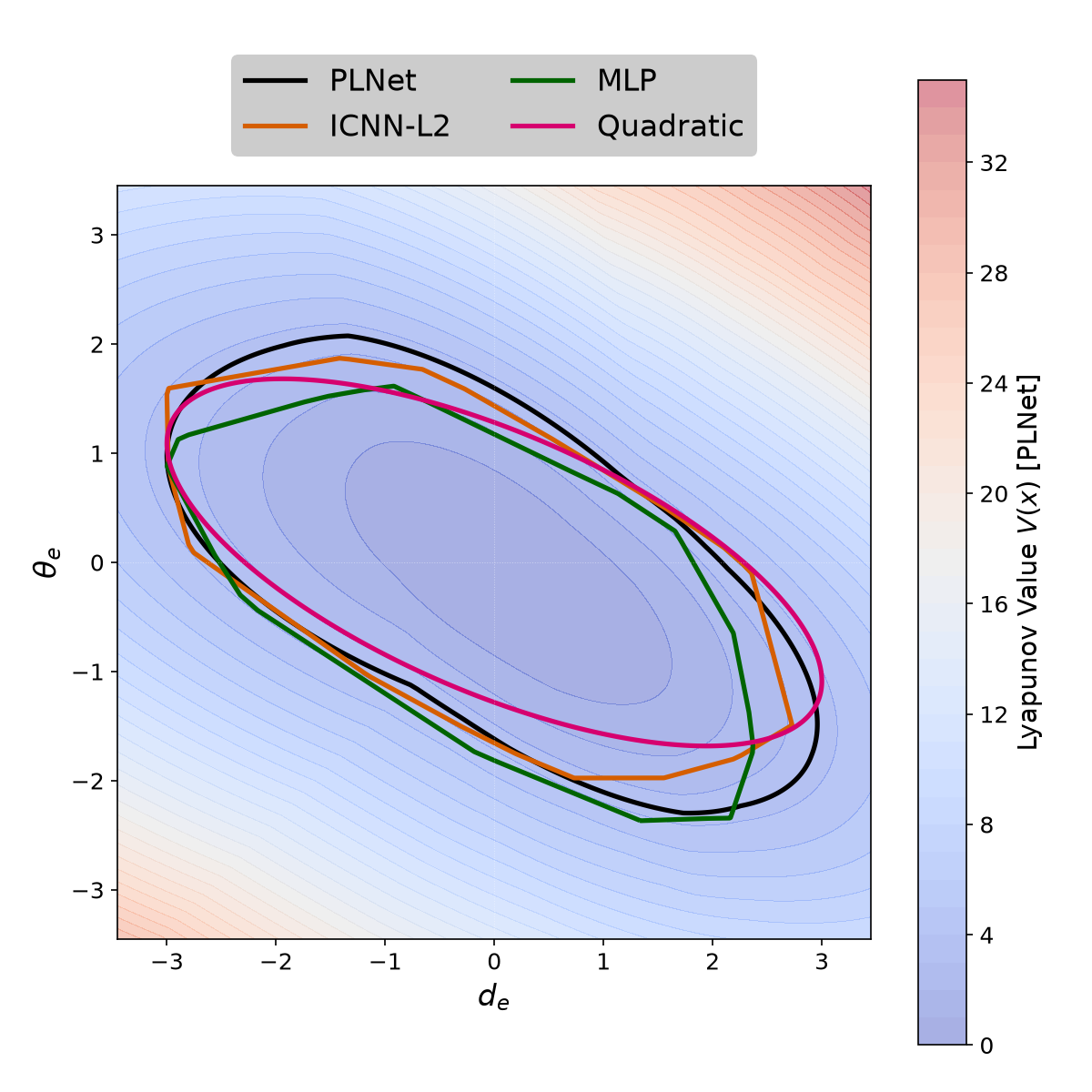}
        \caption{Baseline ROA sublevel sets ($V(x) \leq c_{\text{train}}$).}
        \label{fig:PTbase}
    \end{subfigure}
    \hfill
    \begin{subfigure}[b]{0.23\textwidth}
        \centering
        \includegraphics[width=\textwidth]{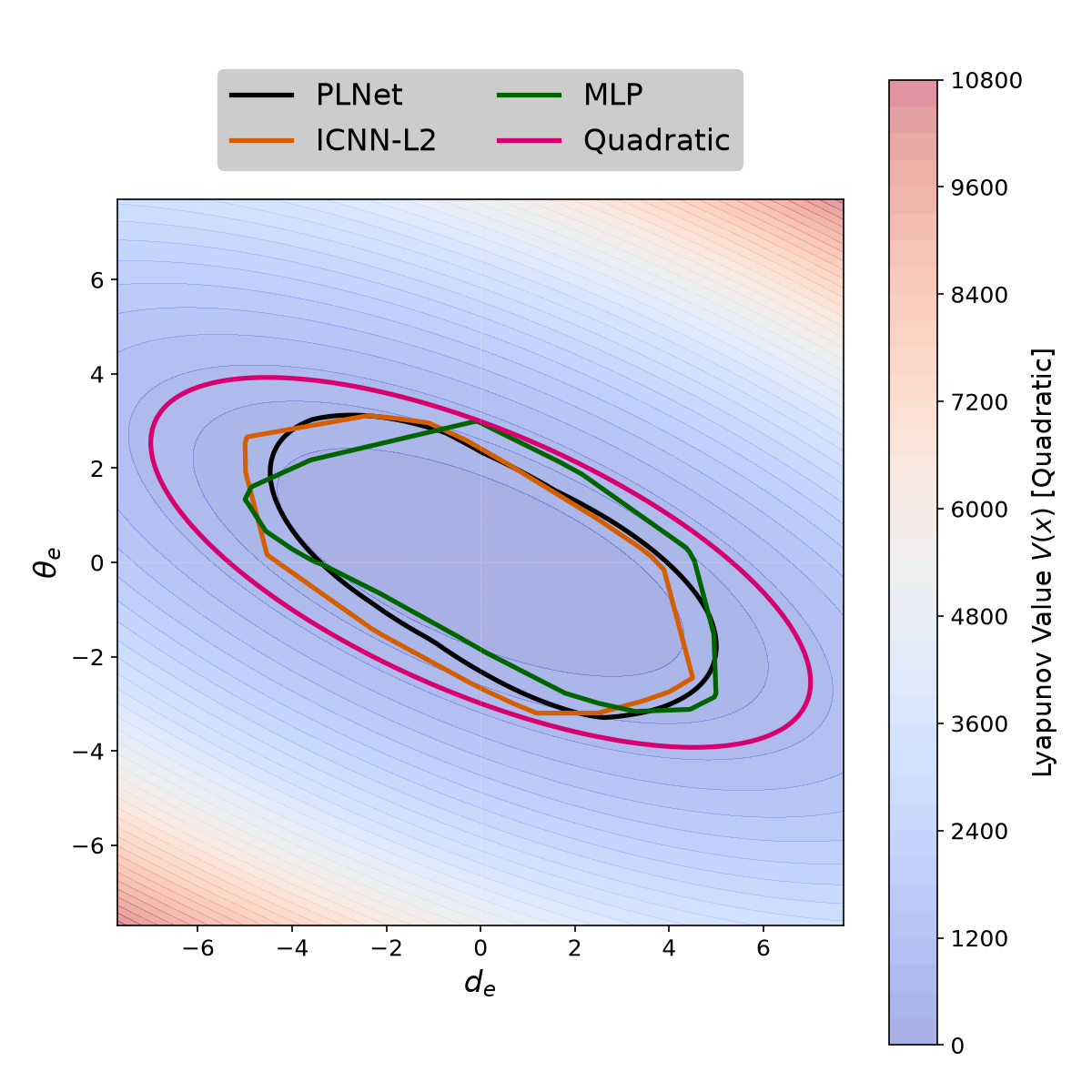}
        \caption{Expanded ROA sublevel sets ($V(x) \leq c_{\text{expanded}}$).}
        \label{fig:PTexpanded}
    \end{subfigure}
    \caption{Comparison of certified ROA for the path tracking system across PLNet, ICNN-L2, MLP and quadratic CLFs, showing the best performing seed for each.}
    \label{fig:ArchPT}
\end{figure}

The results for both benchmarks are summarized in
Tables~\ref{tab:InvertedPendulumResults} and
\ref{tab:PathTrackingResults}. Figures~\ref{fig:ArchIP}
and \ref{fig:ArchPT} compare the base and expanded ROA
sublevel sets for the best-performing seeds of the PLNet,
ICNN-L2, MLP, and quadratic CLF architectures.

\subsubsection{Inverted Pendulum}

\begin{table*}[h]
    \footnotesize
    \centering
    \renewcommand{\arraystretch}{1.6}
    \resizebox{\textwidth}{!}{%
    \begin{tabular}{c|c|c|c|c|c|c|c|c|c|c|c}
    \textbf{Architecture} & \textbf{Base ROA}  & \textbf{Max base ROA}   &  \textbf{Exp. dia.} & \textbf{Exp. ROA} & \textbf{Max exp. ROA} &  \textbf{Phase 2 time (s)} & \textbf{Phase 3 time (s)} & \textbf{Expand time (s)} & \textbf{Total time (s)} & \textbf{Num. rounds} & \textbf{Success \%}\\
    \hline
    ICNN-L2 & $33.632 \pm 21.733$  & $53.649$        & $23.200 \pm 13.084$ & $211.504 \pm 140.483$ & $343.526$            & $9.176 \pm 0.100$  & $186.666 \pm 73.002$ & $134.904 \pm 11.572$ & $358.560 \pm 76.747$ & $5.250 \pm 2.062$ &        80 \\
 ICNN-L1         & $50.404 \pm 18.370$  & $77.624$        & $25.400 \pm 4.336$  & $255.985 \pm 173.185$ & $486.940$            & $9.174 \pm 0.160$  & $92.715 \pm 91.350$  & $109.192 \pm 29.918$ & $239.427 \pm 94.357$ & $5.400 \pm 5.320$ &       100 \\
 MLP             & $40.110 \pm 15.697$  & $60.979$        & $26.200 \pm 7.950$  & $193.872 \pm 123.637$ & $366.314$            & $7.392 \pm 0.124$  & $93.202 \pm 44.626$  & $114.561 \pm 59.097$ & $242.619 \pm 58.066$ & $5.400 \pm 2.702$ &       100 \\
 MLP-ReLU-L1     & $24.540 \pm 23.614$  & $63.480$        & $22.400 \pm 5.683$  & $112.498 \pm 137.071$ & $351.403$            & $5.755 \pm 0.074$  & $152.478 \pm 35.694$ & $88.351 \pm 43.748$  & $274.345 \pm 79.059$ & $9.400 \pm 2.074$ &       100 \\
 PLNet           & $132.273 \pm 11.245$ & $151.771$       & $28.200 \pm 4.025$  & $820.677 \pm 251.977$ & $1040.044$           & $36.468 \pm 0.196$ & $57.348 \pm 10.054$  & $137.452 \pm 30.799$ & $259.062 \pm 33.314$ & $1.000 \pm 0.000$ &       100 \\
 Quadratic       & $60.451 \pm 0.000$   & $60.451$        & $27.200 \pm 6.261$  & $323.751 \pm 120.905$ & $377.822$            &   \multicolumn{2}{c|}{$11.318 \pm 1.035$} & $120.031 \pm 45.109$ & $158.563 \pm 44.578$ & $1.000 \pm 0.000$ &       100 \\
    
    \end{tabular}%
    }
    \caption{Inverted Pendulum Results}
    \label{tab:InvertedPendulumResults}
\end{table*}

The PLNet achieves the largest certified ROAs among
the considered architectures. Its maximum base and
expanded ROAs are $1.955$ and $2.136$ times those of
the second-best architecture (ICNN-L1), respectively.
Moreover, the PLNet exhibits relatively low variability
in its base ROA and requires only one certification
round across all seeds, as does the quadratic CLF. The PLNet also achieves the largest ROAs on average, with $2.188$ and $2.535$ times the second largest (quadratic) average base and expanded ROAs, respectively.
ICNN-L2 is the only architecture that fails to
certify the training region for one seed.


\subsubsection{Path Tracking}
\begin{table*}[h]
    \footnotesize
    \centering
    \renewcommand{\arraystretch}{1.6}
    \resizebox{\textwidth}{!}{%
    \begin{tabular}{c|c|c|c|c|c|c|c|c|c|c|c|c}
    \textbf{Architecture} & \textbf{Base ROA}  & \textbf{Max base ROA}   &  \textbf{Exp. dia.} & \textbf{Exp. ROA} & \textbf{Max exp. ROA} & \textbf{Phase 2 time (s)} & \textbf{Phase 3 time (s)} & \textbf{Expand time (s)} & \textbf{Total time (s)} & \textbf{Num. rounds}  & \textbf{Success \%}\\
    \hline
    ICNN-L2 & $9.281 \pm 2.891$  & $13.818$        & $3.800 \pm 1.095$ & $16.221 \pm 12.393$ & $37.447$             & $9.365 \pm 1.113$  & $101.629 \pm 126.139$ & $15.574 \pm 8.687$  & $155.974 \pm 120.167$ & $4.200 \pm 5.630$ &       100 \\
    ICNN-L1         & $8.733 \pm 3.016$  & $11.899$        & $3.600 \pm 0.894$ & $13.010 \pm 6.797$  & $20.000$             & $8.763 \pm 0.075$  & $56.254 \pm 34.494$   & $13.557 \pm 6.986$  & $106.773 \pm 39.373$  & $3.400 \pm 2.074$ &       100 \\
    MLP             & $9.848 \pm 3.570$  & $12.641$        & $3.800 \pm 1.095$ & $15.953 \pm 11.123$ & $35.358$             & $7.083 \pm 0.057$  & $85.298 \pm 80.465$   & $14.523 \pm 8.364$  & $134.530 \pm 88.341$  & $5.400 \pm 4.930$ &       100 \\
    MLP-ReLU-L1     & $7.432 \pm 3.561$  & $12.844$        & $3.200 \pm 0.447$ & $8.573 \pm 4.230$   & $12.844$             & $5.539 \pm 0.066$  & $98.211 \pm 88.728$   & $9.997 \pm 3.281$   & $141.457 \pm 88.658$  & $7.200 \pm 6.573$ &       100 \\
    PLNet           & $13.634 \pm 0.933$ & $15.055$        & $4.000 \pm 0.707$ & $24.223 \pm 8.125$  & $35.461$             & $35.091 \pm 0.210$ & $71.413 \pm 29.695$   & $22.207 \pm 8.659$  & $156.445 \pm 29.920$  & $1.400 \pm 0.548$ &       100 \\
    Quadratic       & $12.071 \pm 0.000$ & $12.071$        & $6.400 \pm 0.894$ & $55.797 \pm 14.558$ & $65.722$             &  \multicolumn{2}{c|}{$10.638 \pm 0.771$}     & $35.032 \pm 6.910$  & $72.771 \pm 6.863$    & $1.000 \pm 0.000$ &       100 \\
    
    \end{tabular}%
    }
    \caption{Path Tracking Results}
    \label{tab:PathTrackingResults}
\end{table*}

The PLNet achieves the largest maximum base ROA,
exceeding the second-best architecture (ICNN-L2)
by a factor of $1.090$. However, the quadratic CLF
achieves the largest expanded ROA, with a maximum
$1.755$ times that of ICNN-L2, and the largest mean
expanded domain size. This suggests the quadratic CLF generalizes better to regions outside the training region for this system. It also requires only one certification round across all seeds, and all architectures achieve a 100\% success rate. The PLNet achieves the largest base ROA on average, being $1.129$ times larger than the second-best architecture (quadratic). The quadratic CLF achieves the largest expanded ROA on average, being $2.303$ times the second-best architecture (PLNet). Considering the neural network CLFs, the PLNet achieves the largest expanded ROA on average, being $1.493$ times larger than the second-best neural CLF architecture (ICNN-L2), and achieves the largest mean expanded domain size.

\subsection{Influence of PLNet Lipschitz Bounds} \label{sec:shape_ROA}
\begin{figure}[h]
    \centering
    \begin{subfigure}[b]{0.23\textwidth}
        \centering
        \includegraphics[width=\textwidth]{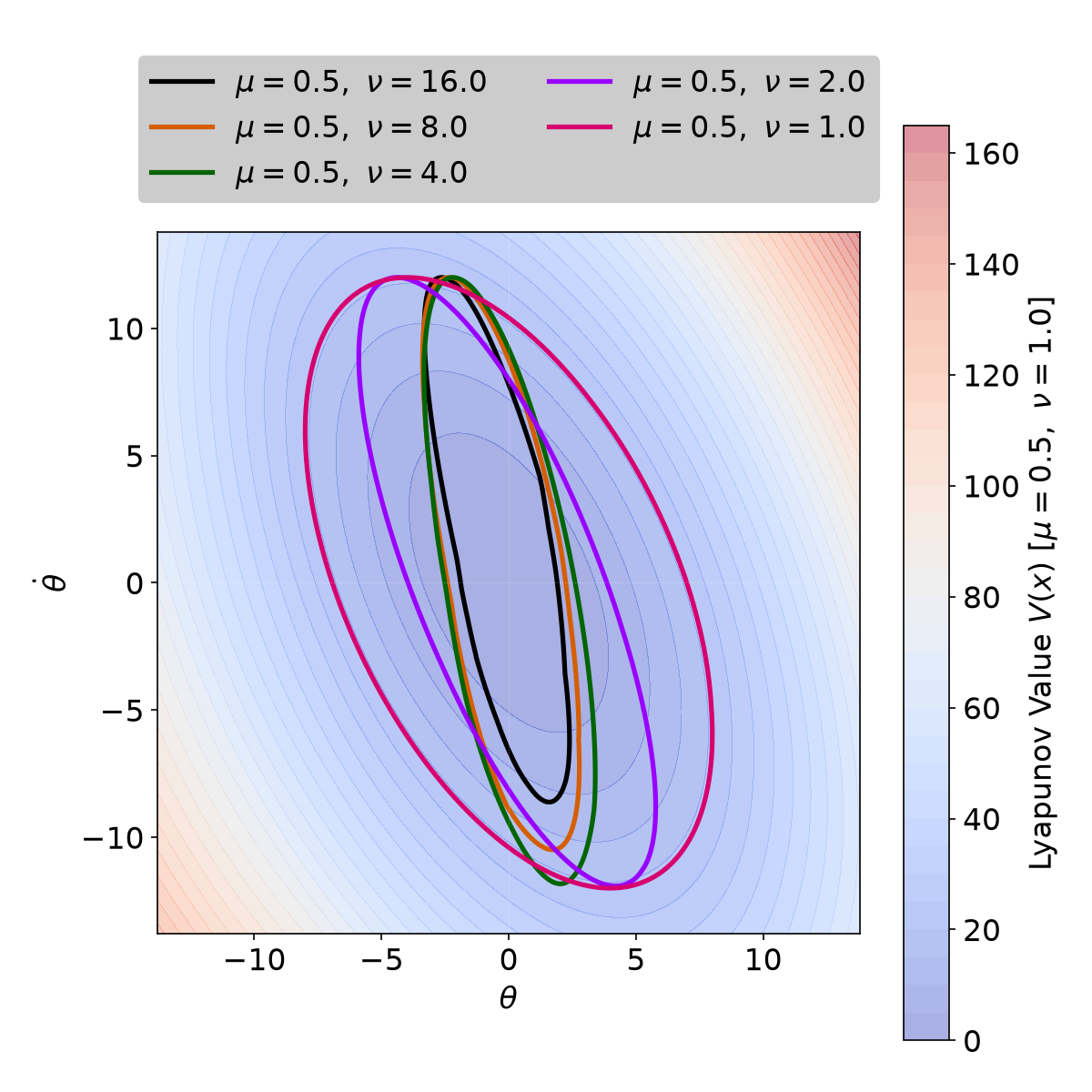}
        \caption{Baseline ROA sublevel sets ($V(x) \leq c_{\text{train}}$).}
        \label{fig:munubase}
    \end{subfigure}
    \hfill
    \begin{subfigure}[b]{0.23\textwidth}
        \centering
        \includegraphics[width=\textwidth]{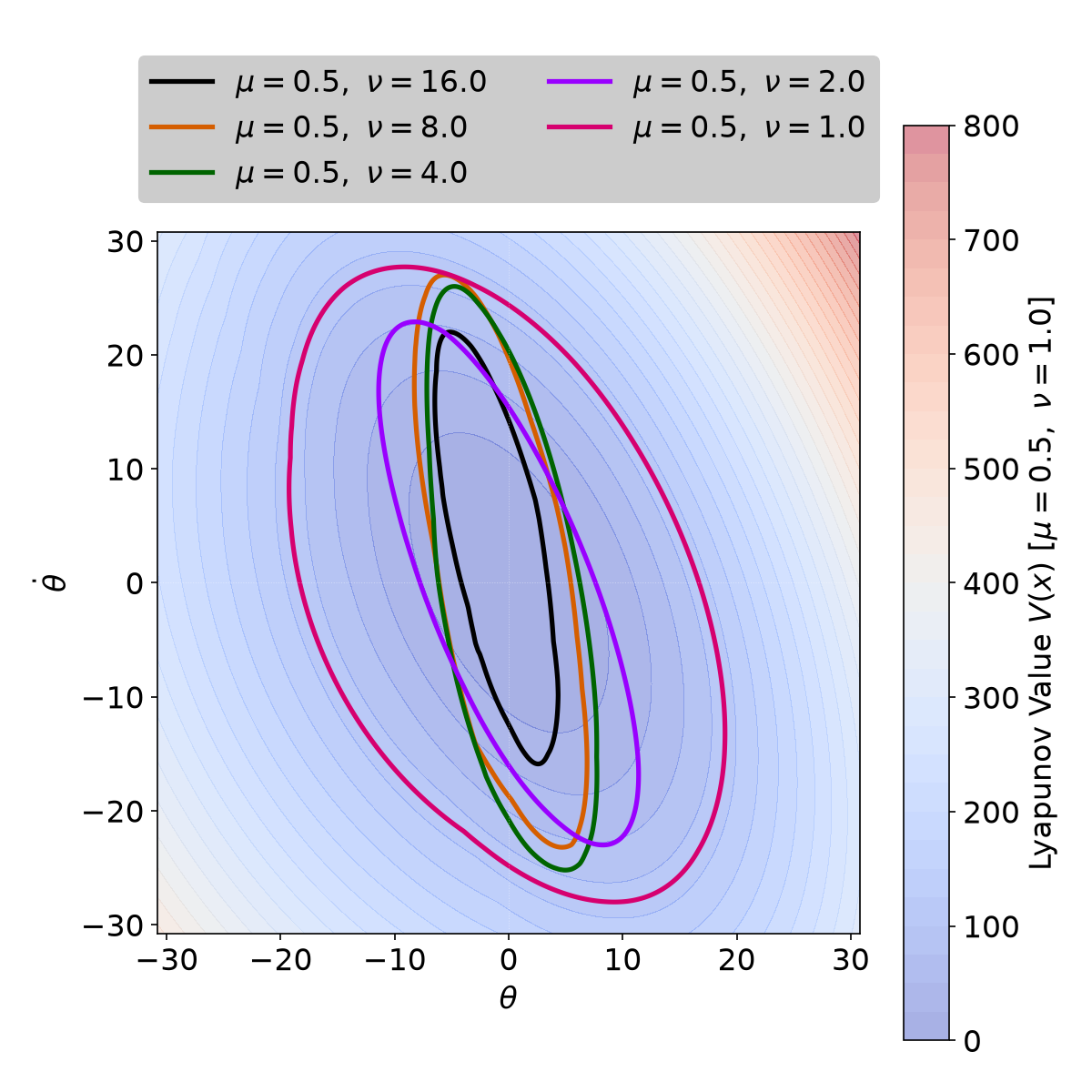}
        \caption{Expanded ROA sublevel sets ($V(x) \leq c_{\text{expanded}}$).}
        \label{fig:munuexpanded}
    \end{subfigure}
    \caption{Effect of the Lipschitz bound $\nu$ on the certified region of attraction for the inverted pendulum, with $\mu = 0.5$ fixed.}
    \label{fig:IPmunucomparison}
\end{figure}

\begin{figure}[h]
    \centering
    \begin{subfigure}[b]{0.23\textwidth}
        \centering
        \includegraphics[width=\textwidth]{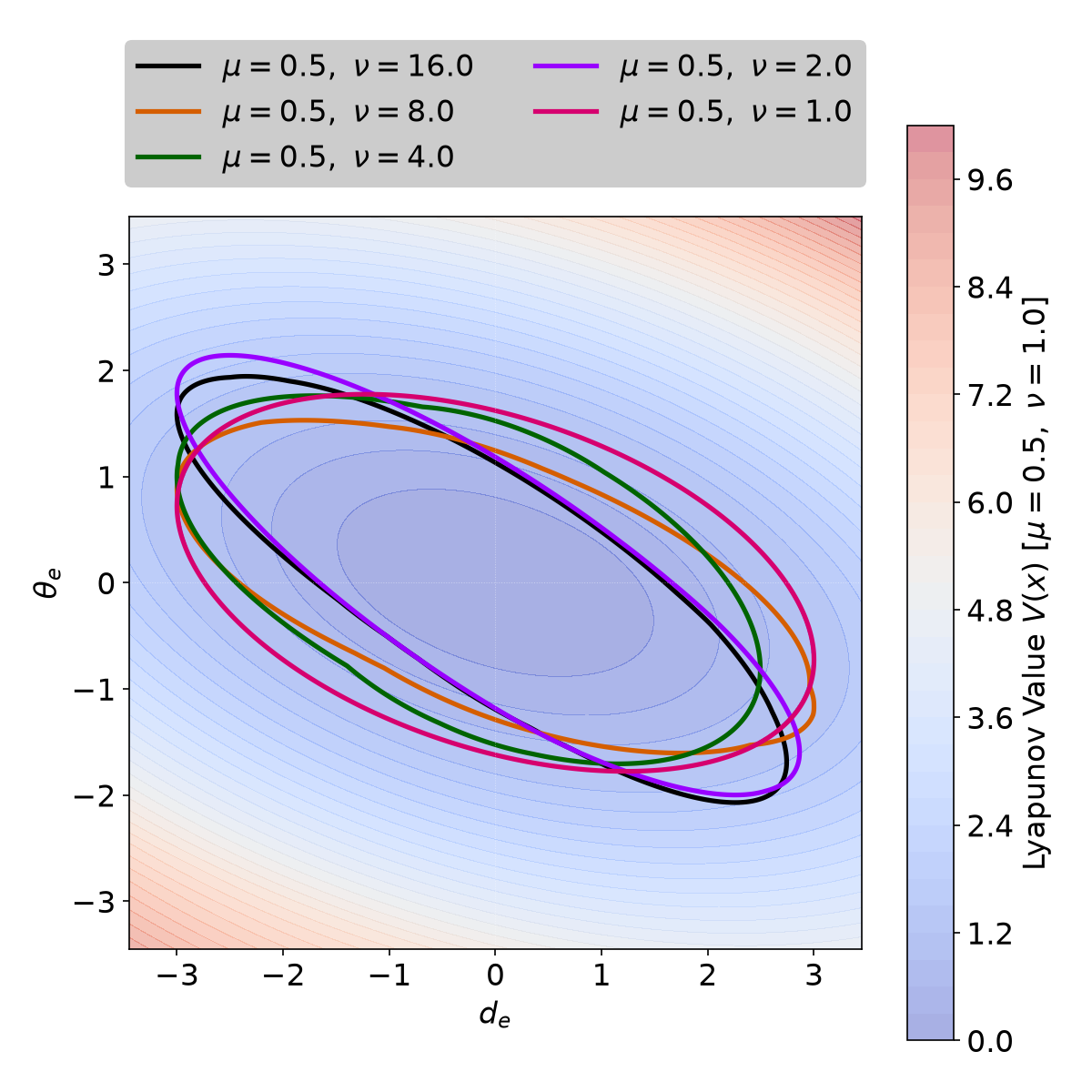}
        \caption{Baseline ROA sublevel sets ($V(x) \leq c_{\text{train}}$).}
        \label{fig:munubase_PT}
    \end{subfigure}
    \hfill
    \begin{subfigure}[b]{0.23\textwidth}
        \centering
        \includegraphics[width=\textwidth]{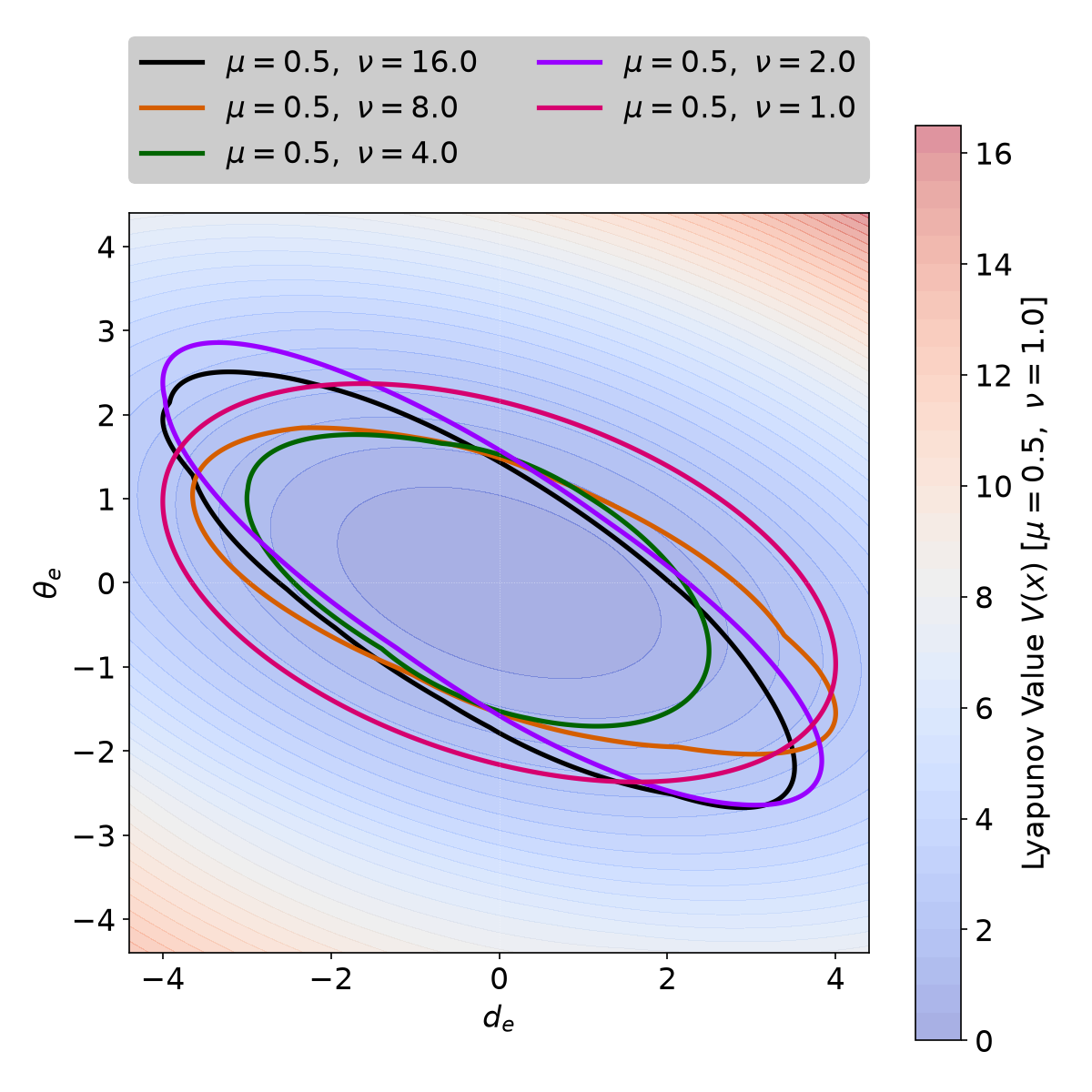}
        \caption{Expanded ROA sublevel sets ($V(x) \leq c_{\text{expanded}}$).}
        \label{fig:munuexpanded_PT}
    \end{subfigure}
    \caption{Effect of the Lipschitz bound $\nu$ on the certified region of attraction for the path tracking system, with $\mu = 0.5$ fixed.}
    \label{fig:PTmunucomparison}
\end{figure}

To investigate the influence of the PLNet's monotonicity
constant $\mu$ and Lipschitz constant $\nu$, introduced
in Section~\ref{sec:plnet}, we fix $\mu=0.5$ and vary
$\nu$ for both systems, using seed 10 without a time
limit. The results in Tables~\ref{tab:InvertedPendulumPLNetmunu}
and \ref{tab:PathTrackingPLNetmunu} and
Figures~\ref{fig:IPmunucomparison} and
\ref{fig:PTmunucomparison} show that smaller ratios
$\nu : \mu$ generally yield more isotropic sublevel
sets and can increase the certified ROA. The smallest
tested ratio achieves the largest ROA for both systems,
with areas of 1511 and 27.14 for the inverted pendulum
and path-tracking system, respectively. However,
overly restrictive bounds may limit the network's
expressivity and prevent successful training and
verification.

\begin{table}[h]
    \footnotesize
    \centering
    \renewcommand{\arraystretch}{1.6}
    \resizebox{0.48\textwidth}{!}{%
    \begin{tabular}{c|c|c|c|c|c|c|c}
    \textbf{$(\mu, \nu)$} & \textbf{Exp. dia.}  & \textbf{Base ROA} & \textbf{Exp. ROA} & \textbf{Phase 3 time (s)} & \textbf{Expand time (s)} & \textbf{Total time (s)} & \textbf{Num. rounds}\\
    \hline
    \textbf{(0.5, 1.0)} & 28.00 & 261.807 & 1511.088 & 165.725 & 150.853 & 384.931 & 3\\
    \textbf{(0.5, 2.0)} & 23.00 & 147.217 & 557.744 & 73.587 & 105.125 & 246.713 & 1\\
    \textbf{(0.5, 4.0)} & 26.00 & 97.316 & 493.859 & 100.843 & 141.190 & 315.105 & 1\\
    \textbf{(0.5, 8.0)} & 27.00 & 84.387 & 463.968 & 151.987 & 153.294 & 378.748 & 1\\
    \textbf{(0.5, 16.0)} & 22.00 & 63.597 & 231.909 & 161.718 & 103.479 & 329.889 & 1
    \end{tabular}%
    }
    \caption{Inverted Pendulum PLNet $\mu$ and $\nu$ variation.}
    \label{tab:InvertedPendulumPLNetmunu}
\end{table}

\begin{table}[h]
    \footnotesize
    \centering
    \renewcommand{\arraystretch}{1.6}
    \resizebox{0.48\textwidth}{!}{%
    \begin{tabular}{c|c|c|c|c|c|c|c}
    \textbf{$(\mu, \nu)$} & \textbf{Exp. dia.}  & \textbf{Base ROA} & \textbf{Exp. ROA} & \textbf{Phase 3 time (s)} & \textbf{Expand time (s)} & \textbf{Total time (s)} & \textbf{Num. rounds}\\
    \hline
    \textbf{(0.5, 1.0)} & 4.00 & 15.264 & 27.137 & 44.510 & 25.194 & 131.684 & 1\\
    \textbf{(0.5, 2.0)} & 4.00 & 10.865 & 19.532 & 46.272 & 19.193 & 127.167 & 1\\
    \textbf{(0.5, 4.0)} & 3.00 & 12.808 & 12.808 & 46.536 & 10.428 & 119.453 & 1\\
    \textbf{(0.5, 8.0)} & 4.00 & 11.718 & 17.686 & 50.093 & 23.492 & 146.982 & 1\\
    \textbf{(0.5, 16.0)} & 4.00 & 10.437 & 18.805 & 866.815 & 25.472 & 956.050 & 2
    \end{tabular}%
    }
    \caption{PathTracking PLNet $\mu$ and $\nu$ variation.}
    \label{tab:PathTrackingPLNetmunu}
\end{table}

\section{Conclusions}\label{sec:conclusions}

This work developed a framework for jointly learning neural controllers and control Lyapunov functions (CLFs) for nonlinear discrete-time systems. By exploiting structured neural architectures, particularly ICNNs and PLNets, positive definiteness of the CLF candidates is guaranteed by construction, reducing the conditions that require formal verification. To accommodate these architectures, we extended mixed-integer verification beyond conventional MILP formulations to MIQP formulations, enabling the verification of quadratic neural Lyapunov functions. Furthermore, we introduced an iterative ROA expansion procedure to certify larger regions beyond the initial training domain without retraining.

Numerical experiments demonstrated the advantages of structured neural CLFs in obtaining larger certified regions and showed how the monotonicity and Lipschitz bounds of PLNets influence their geometry. Future work will investigate extensions to higher-dimensional systems and more sophisticated ROA expansion strategies.

\section*{Acknowledgments}
ChatGPT (OpenAI, GPT-6) was used to assist with language editing and improving the clarity and conciseness of the manuscript. Claude (Anthropic) and Gemini (Google) were used to support code development and debugging. The authors reviewed and verified all AI-assisted contributions and take full responsibility for the scientific content and results.

\bibliographystyle{IEEEtran}   
\bibliography{references}

@article{chang2019neural,
  title={Neural {L}yapunov control},
  author={Chang, Ya-Chien and Roohi, Nima and Gao, Sicun},
  journal={Advances in Neural Information Processing Systems},
  volume={32},
  year={2019}
}

@article{wu2023neural,
  title={Neural {L}yapunov control for discrete-time systems},
  author={Wu, Junlin and Clark, Andrew and Kantaros, Yiannis and Vorobeychik, Yevgeniy},
  journal={Advances in Neural Information Processing Systems},
  volume={36},
  pages={2939--2955},
  year={2023}
}

@inproceedings{gowal2019scalable,
  title     = {Scalable Verified Training for Provably
               Robust Image Classification},
  author    = {Gowal, Sven and Dvijotham, Krishnamurthy
               and Stanforth, Robert and Bunel, Rudy
               and Qin, Chongli and Uesato, Jonathan
               and Arandjelovic, Relja and Mann, Timothy
               and Kohli, Pushmeet},
  booktitle = {IEEE/CVF International
               Conference on Computer Vision},
  pages     = {4842--4851},
  year      = {2019},
  doi       = {10.1109/ICCV.2019.00494}
}

@inproceedings{amos2017inputconvexneuralnetworks,
  title     = {Input Convex Neural Networks},
  author    = {Amos, Brandon and Xu, Lei and
               Kolter, J. Zico},
  booktitle = {International
               Conference on Machine Learning},
  volume    = {70},
  pages     = {146--155},
  year      = {2017},
  publisher = {PMLR}
}

@inproceedings{wang2024monotone,
  title     = {Monotone, Bi-Lipschitz, and
               Polyak-{\L}ojasiewicz Networks},
  author    = {Wang, Ruigang and
               Dvijotham, Krishnamurthy Dj and
               Manchester, Ian},
  booktitle = {International
               Conference on Machine Learning},
  volume    = {235},
  pages     = {50379--50399},
  year      = {2024},
  publisher = {PMLR}
}

@inproceedings{cheng2024learningstablepassiveneural,
  title     = {Learning Stable and Passive Neural
               Differential Equations},
  author    = {Cheng, Jing and Wang, Ruigang and
               Manchester, Ian R.},
  booktitle = {Conference on
               Decision and Control},
  pages     = {7859--7864},
  year      = {2024},
  publisher = {IEEE},
  doi       = {10.1109/CDC56724.2024.10886407}
}

@article{kolter2019learning,
  title={Learning stable deep dynamics models},
  author={Kolter, J Zico and Manek, Gaurav},
  journal={Advances in neural information processing systems},
  volume={32},
  year={2019}
}

@inproceedings{Mukherjee2022,
  title={Neural {L}yapunov differentiable predictive control},
  author={Mukherjee, Sayak and Drgo{\v{n}}a, J{\'a}n and Tuor, Aaron and Halappanavar, Mahantesh and Vrabie, Draguna},
  booktitle={Conference on Decision and Control},
  pages={2097--2104},
  year={2022},
  organization={IEEE}
}

@inproceedings{xu2020automatic,
  title     = {Automatic Perturbation Analysis for
               Scalable Certified Robustness and Beyond},
  author    = {Xu, Kaidi and Shi, Zhouxing and
               Zhang, Huan and Wang, Yihan and
               Chang, Kai-Wei and Huang, Minlie and
               Kailkhura, Bhavya and Lin, Xue and
               Hsieh, Cho-Jui},
  booktitle = {Advances in Neural Information
               Processing Systems},
  volume    = {33},
  year      = {2020},
  publisher = {Curran Associates, Inc.}
}

@article{Snider2009AutomaticSM,
  title={Automatic steering methods for autonomous automobile path tracking},
  author={Snider, Jarrod M and others},
  journal={Robotics Institute, Pittsburgh, PA, Tech. Rep. CMU-RITR-09-08},
  year={2009}
}

@inproceedings{dai2021lyapunov,
  title     = {Lyapunov-stable neural-network control},
  author    = {Dai, Hongkai and Landry, Benoit and
               Yang, Lujie and Pavone, Marco and
               Tedrake, Russ},
  booktitle = {Robotics: Science and Systems},
  year      = {2021},
  doi       = {10.15607/RSS.2021.XVII.063}
}

@article{li2026two,
  title={Two-Stage Learning of Stabilizing Neural Controllers via Zubov Sampling and Iterative Domain Expansion},
  author={Li, Haoyu and Zhong, Xiangru and Hu, Bin and Zhang, Huan},
  journal={Advances in Neural Information Processing Systems},
  volume={38},
  pages={73980--74009},
  year={2026}
}

@inproceedings{liu2025formally,
  title={Formally verified physics-informed neural control lyapunov functions},
  author={Liu, Jun and Fitzsimmons, Maxwell and Zhou, Ruikun and Meng, Yiming},
  booktitle={2025 American Control Conference (ACC)},
  pages={1347--1354},
  year={2025},
  organization={IEEE}
}

@article{zhou2022neural,
  title={Neural {L}yapunov control of unknown nonlinear systems with stability guarantees},
  author={Zhou, Ruikun and Quartz, Thanin and De Sterck, Hans and Liu, Jun},
  journal={Advances in Neural Information Processing Systems},
  volume={35},
  pages={29113--29125},
  year={2022}
}

@inproceedings{yang2024lyapunov,
  title     = {Lyapunov-stable Neural Control for
               State and Output Feedback: A Novel Formulation},
  author    = {Yang, Lujie and Dai, Hongkai and
               Shi, Zhouxing and Hsieh, Cho-Jui and
               Tedrake, Russ and Zhang, Huan},
  booktitle = {International
               Conference on Machine Learning},
  volume    = {235},
  pages     = {56033--56046},
  year      = {2024},
  publisher = {PMLR}
}

@inproceedings{shi2026certified,
  title     = {Certified Training with Branch-and-Bound
               for Lyapunov-stable Neural Control},
  author    = {Shi, Zhouxing and Li, Haoyu and
               Hsieh, Cho-Jui and Zhang, Huan},
  booktitle = {Annual Learning
               for Dynamics and Control Conference},
  series    = {Proceedings of Machine Learning Research},
  volume    = {331},
  pages     = {788--808},
  year      = {2026},
  publisher = {PMLR}
}

@inproceedings{min2023data,
  title={Data-driven control with inherent {L}yapunov stability},
  author={Min, Youngjae and Richards, Spencer M and Azizan, Navid},
  booktitle={Conference on Decision and Control},
  pages={6032--6037},
  year={2023},
  organization={IEEE}
}

@misc{IBMCPLEX,
  author       = {{IBM}},
  title        = {{IBM ILOG CPLEX Optimization Studio}},
  howpublished = {\url{https://www.ibm.com/products/ilog-cplex-optimization-studio}},
  year         = {2026},
  note         = {Software}
}

\end{document}